\documentclass[a4paper, 12pt, openany, oneside]{article}
\usepackage[cp1251]{inputenc}
\usepackage[english, russian]{babel}
\usepackage{amsmath, latexsym,  amssymb, bm,  graphics, eucal, amsfonts, array}
\usepackage[dvips]{graphicx}
\makeatletter

\newcommand{\rot}{\mathop{\rm rot\,}}
\makeatother
\usepackage{indentfirst}
\makeatother

\begin{document}
\renewcommand{\refname}{\begin{center} \bf References\end{center}}
\thispagestyle{empty}
\renewcommand{\abstractname}{Abstract}
\renewcommand{\contentsname}{Contents}
\large
\newcommand{\mc}[1]{\mathcal{#1}}
\newcommand{\E}{\mc{E}}

 \begin{center}
\bf The occurrence of transverse and longitudinal electric currents in the classical
plasma under the action of $N$  transverse electromagnetic waves
\end{center}\bigskip

\begin{center}
  \bf A. V. Latyshev\footnote{avlatyshev$@$mail.ru} and
  V. I. Askerova\footnote{vera$ \_ $askerova$@$mail.ru}
\end{center}\medskip
\bigskip

\begin{center}
{\it Faculty of Physics and Mathematics,\\ Moscow State Regional
University, 105005,\\ Moscow, Radio str., 10-A}
\end{center}\medskip

\begin{abstract}
Classical plasma with arbitrary degree of degeneration of electronic gas is consi\-dered.
In plasma $N$ ($N>2$) collinear electromagnatic waves are propagated. It is required to
find the response of plasma to these waves. Distribution function  in square-law
approximation on quantities of two small parameters from Vlasov equation is received.
The formula for electric current calculation is deduced.
It is demonstrated that the nonlinearity account leads to
occurrence of the longitudinal electric current directed along a  wave  vector. This
longitudinal current is orthogonal to the  known transversal current received at the linear
analysis. The case of small values of wave number is considered.

{\bf{Key words}:} Vlasov equation, classical plasma, transversal
 and longitudinal and transversal electric currents, nonlinear analysis.

PACS numbers:  05.20.Dd Kinetic theory, 52.25.Dg Plasma kinetic equations.
\end{abstract}

\begin{center}
\section{Introduction}
\end{center}

In the present work formulas are deduced for electric current calculation in
classical collisionless plasma. At the solution of the kinetic Vlasov equation
describing behaviour of classical degenerate plasmas, we consider as in
 $\qquad$ decomposition distribution functions, and in decomposition of quantity of the
self-conjugate electromagnetic field the quantities proportional to square of
intensity of an external electric field. In such nonlinear approach it appears that
the electric current has two nonzero components. One component of an electric
current it is directed along vector potentials of electromagnetic fields. These
components of an electric current precisely same, as well as in the linear analysis.
It is a "transversal"\,
 current.

Those, in linear approach we receive known expression of a transversal electric current.

The second nonzero an electric current component has the second
order of smallness concerning quantities intensity of electric
fields. The second electric current component is directed
along a wave vector. This current is orthogonal to the first
a component. It is a "longitudinal"\, current.

Occurrence of a longitudinal current comes to light the spent
nonlinear analysis of interaction of electromagnetic fields
with plasma.

Nonlinear effects in plasma are studied already long time
\cite{Gins}--\cite{Brod}.

In works \cite{Gins} and \cite{Zyt} nonlinear effects in plasma
are studied. In work \cite{Zyt} nonlinear current was used, in
particular, in probability questions decay processes. We will
notice, that in work \cite{Zyt2} it is underlined existence
of nonlinear current along a wave vector (see the formula
(2.9) from  \cite{Zyt2}).

In experimental work \cite{Akhmediev} the contribution normal
field components in a nonlinear superficial current in a signal
of the second harmonic is found out. In works \cite{Urupin1, Urupin2}
generation of a nonlinear superficial current was studied at
interaction of a laser impulse with metal.

We will specify in a number of works on plasma, including to
the quantum. These are works  \cite{Lat3}--\cite{Lat11}.\\

\section{The Vlasov equation}

Let us demonstrate, that in case of the classical plasma described by
kinetic Vlasov equation, the longitudinal current is generated and we
 will calculate its density. It was specified in existence of this
  current more half a century ago \cite{Gins}.

Let us consider that  the $ N $ electromagnetic waves are propagated with strengths
$$
\textbf{E}_{j}=\textbf{E}_{0j}e^{i\left(\textbf{k}_j\textbf{r}-
    \omega_{j}t \right) },
\qquad \textbf{H}_{j}=\textbf{H}_{0j}e^{i\left( \textbf{k}_j\textbf{r}-
    \omega_{j}t \right) },
$$
where $j=1,2,\cdots,N. $

Let us assume that directions of propagation of waves are collinear, that is $
\textbf{k}_{1}\parallel \textbf{k}_{2}\parallel \cdots
\parallel \textbf{k}_{N}. $

We will consider a case, when the directions
electric (and magnetic) fields of waves are collinear
 $ \textbf{E}_{1}\parallel \textbf{E}_{2}\parallel \cdots \parallel
 \textbf{E}_{N} $, $ \left( \textbf{H}_{1}\parallel \textbf{H}_{2}\parallel
 \cdots \parallel \textbf{H}_{N}\right)$.
Correspon\-ding electric and magnetic fields are connected with vector
  potentials equalities
$$
\textbf{E}_{j}=-\frac{1}{c}\dfrac{\partial \textbf{A}_{j}}{\partial t}=
\dfrac{i\omega_{j}}{c}\textbf{A}_j,
\qquad
\textbf{H}_{j}=\rot\textbf{A}_{j},\qquad
j=1,2,\cdots,N.
$$

We take the Vlasov equation describing behavior of classical collisionless plasma
$$
\frac{\partial f}{\partial t}+\textbf{v}\frac{\partial f}{\partial\textbf{r}}+
e\left( \sum\limits_{j=1}^{N}\textbf{E}_j
+\frac{1}{c}\left[ \textbf{v},\sum\limits_{j=1}^{N}\textbf{H}_j\right]\right)
\frac{\partial f}{\partial\textbf{p}}=0.
\eqno{(\textup{1.1})}
$$

In the equation (1.1) $ f $ is cumulative distribution function of electrons of
plasma, $\textbf{E}_j,\textbf{H}_j (j=1,2,\cdots,N)$ are components of an
electromagnetic field,$ c $ is the velocity of light, $ \textbf{p}=m\textbf{v} $ is
momentum of electrons, ${\bf v}$ is the electrons velocity,
$f^{(0)}=f_{eq}(\textbf{r},v)$ (eq $ \equiv $ equilibrium ) is
local equilibrium distribution of Fermi---Dirac
$$
f_{eq}\left( \textbf{r},v \right)=\dfrac{1}{1+\rm exp\left(\dfrac{\E -
\mu (\textbf{r})}{k_{B}T} \right)},
$$
or, in dimensionless form,
$$
f_{eq}\left( \textbf{r},v \right)=
\dfrac{1}{1+\rm exp \left(P^{2}-\alpha(\textbf{r})\right)}=
 f_{eq}\left(\textbf{r},P \right),
$$

\noindent$ \E=mv^{2}/2 $ is the electron energy, $ \mu $ is  the chemical potential of
 electronic gas, $ k_{B} $ is the Boltzmann constant, $ T $ is the plasma temperature,
 $ \textbf{P}=\textbf{P}/p_{T} $ is dimensionless momentum of the electrons,
  $ p_{T}=mv_{T} $, $v_{T}  $ is the heat electron velocity, $ v_{T}=\sqrt{2k_{B}T/m},
  \alpha=\mu/(k_{B}T) $ is  the chemical potential,
  $ k_{B}T=\E_{T}=mv_{T}^{2}/2 $ is the heat kinetic electron energy.

Lower local equilibrium distribution of Fermi---Dirac is required to us,
 $$
 f_{0}\left( \textbf v \right)=\left[1+\rm exp\left(\dfrac{\E
    -\mu}{k_{B}T} \right)\right]^{-1}=
 \left[1+\rm exp \left(P^{2}-\alpha \right)\right]^{-1}=f_{0}\left(P \right).
 $$

 It is necessary to specify, that vector potential of an electromagnetic field
 ${\bf A}_j({\bf r},t)$  is orthogonal to a wave vector ${\bf
    k}_j$, i.е.
 $$
 {\bf k}_j\cdot {\bf A}_j({\bf r},t)=0,\qquad j=1,2,\cdots,N.
 $$
 It means that the wave vector ${\bf k}_j$ is orthogonal
 to electric and magnetic fields
 $$
 {\bf k}_j\cdot {\bf E}_j({\bf r},t)=
 {\bf k}_j\cdot {\bf H}_j({\bf r},t)=0,\qquad j=1,2,\cdots,N.
 $$
 For definiteness we will consider, that wave vectors N
 of fields are directed along an axis $x$ and electromagnetic fields
 are directed along an axis  $y$, i.e.
 $$
 {\bf k}_j=k_j(1,0,0), \quad {\bf E}_j=E_{j}(x,t)(0,1,0).
 $$

   Therefore
$$
\textbf{E}_{j}=-\dfrac{1}{c}\dfrac{\partial
\textbf{A}_{j}}{\partial t}=\dfrac{i\omega_{j}}{c} \textbf{A}_{j},\quad
\textbf{H}_{j}=\dfrac{ck_{j}}{\omega}E_{j}\cdot\left(0,0,1 \right),
\quad j=1,2,\cdots,N.
   $$

   Let us find a vector product from the equation (1.1)
   $$
   \left[ \textbf{v},\textbf{H}_{j}\right] =\dfrac{ck_{j}}{\omega_{j}}E_{j}
   \left( v_{y},-v_{x},0\right),
   $$
   then
   $$
   \left[\textbf{v},\sum\limits_{j=1}^{N}\textbf{H}_j\right]=
   \sum\limits_{j=1}^{N} \dfrac{ck_{j}}{\omega_{j}}E_{j}\left( v_{y},-v_{x},0\right).
   $$

  We find Lorentz force by means of a vector product
  $$
  e\left(\textbf{E}_{j}+\frac{1}{c}\left[ \textbf{v},\textbf{H}_{j} \right]
    \right)\dfrac{\partial f}{\partial \textbf{p}} =
  $$
  $$
  =\frac{e}{\omega_{j}}E_{j}\left[ k_{j}v_{y}\dfrac{\partial f}{\partial p_{x}}+
  \left( \omega_{j}-k_{j}v_{x}\right) \dfrac{\partial f}{\partial p_{y}}\right],
  \qquad (j=1,2,\cdots,N).
  $$

  Let us notice that
  $$
  \left[ \textbf{v},\textbf{H}_{j} \right]\dfrac{\partial f_{0}}{\partial\textbf{p}}=0,
  $$
  as
    $$
    \dfrac{\partial f_{0}}{\partial\textbf{p}}\sim \textbf{v}.
    $$

    Now the equation (1.1) is somewhat simplified:

    $$
    \frac{\partial f}{\partial t}+v_{x}\frac{\partial f}{\partial x}+
    e \sum\limits_{j=1}^{N}\frac{E_{j}}{\omega_{j}}
    \left[ k_{j}v_{y}\dfrac{\partial f}{\partial p_{x}}+
    \left( \omega_{j}-k_{j}v_{x}\right)
    \dfrac{\partial f}{\partial p_{y}}\right]=0.
\eqno{(\textup{1.2})}
    $$

    We will search the solution of equation (1.2) in the form
    $$
    f=f_{0}\left(P\right)+f_{1}+f_{2}.\eqno{(\textup{1.3})}
    $$
    Here
    $$
    f_{1}=E_{1}\varphi_{1}+E_{2}\varphi_{2}+\cdots+
    E_{N}\varphi_{N}=
    \sum\limits_{j=1}^{N}E_{j}\varphi_{j},\eqno{(\textup{1.4})}
    $$
    where
    $$
    E_{j}\sim e^{i(k_{j}x-\omega t) },
    $$
    and
    $$
    f_{2}=\sum\limits_{j=1}^{N}E_{j}^{2}\psi_{j}+
    \sum\limits_{\substack{b,s=1\\b<s}}^{N}E_{b}E_{s}\xi_{b,s},
    \eqno{(\textup{1.5})}
    $$
    where
$$
E^{2}_{j}\sim e^{2i(k_{j}x-\omega_{j} t) },
$$
$$
E_{b}E_{s}\sim e^{i[(k_{b}+k_{s})x-(\omega_{b}+\omega_{s})t]}.
$$

\begin{center}
\section{The solution of Vlasov equation in first approximation }
\end{center}

In this equation exist $2N$ parameters of dimension of length
$\lambda_j={v_T}/{\omega_j}$ ($v_T$ is the heat electron velocity)
and $l_j={1}/{k_j}$.
We shall believe, that on lengths $\lambda_j$,so and on lengths $l_j$
energy variable of electrons under acting correspond electric field $A_j$
 is much less than heat energy of electrons
$k_BT$ ($k_B$ is  Boltzmann constant, $T$ is temperature of plasma ),
i.e. we shall consider small parameters
$$
\alpha_j=\dfrac{{\left|eA_j\right|v_T}}{c k_BT}\qquad (j=1,2,\cdots,N)
$$
and
$$
\beta_j=\dfrac{{\left|eA_j\right|\omega_j}}{k_j k_BTc}\qquad (j=1,2,\cdots,N).
$$

If to use communication of vector potentials
electromagnetic fields with strengths of corresponding
electric fields, injected small parameters are expressed
following equalities
$$
\alpha_j=\dfrac{{\left|eE_j\right|v_T}}{\omega_j k_BT}\qquad (j=1,2,\cdots,N)
$$
and
$$
\beta_j=\dfrac{{\left|eE_j\right|}}{k_j k_BT}\qquad (j=1,2,\cdots,N).
$$

We will work with a method consecutive approximations, considering, that
$$
\alpha_j\ll 1 \qquad (j=1,2,\cdots,N)
$$
and
$$
\beta_j\ll 1  \qquad (j=1,2,\cdots,N).
$$
The equation (1.2) by means of (1.3) is equivalent to the following equations

$$
\frac{\partial f_{1}}{\partial t}+v_{x}\frac{\partial f_{1}}{\partial x}=
-e \sum\limits_{j=1}^{N}\frac{E_{j}}{\omega_{j}}
\left[ k_{j}v_{y}\dfrac{\partial f_{0}}{\partial p_{x}}+
\left( \omega_{j}-k_{j}v_{x}\right)
\dfrac{\partial f_{0}}{\partial p_{y}}\right]\eqno{(\textup{2.1})}
$$
and
$$
\frac{\partial f_{2}}{\partial t}+v_{x}\frac{\partial f_{2}}{\partial x}=
-e \sum\limits_{j=1}^{N}\frac{E_{j}}{\omega_{j}}
\left[ k_{j}v_{y}\dfrac{\partial f_{1}}{\partial p_{x}}+
\left( \omega_{j}-k_{j}v_{x}\right)
\dfrac{\partial f_{1}}{\partial p_{y}}\right].\eqno{(\textup{2.2})}
$$

In the first approximation we search the solution of Vlasov
equation in the form
$$
f=f^{(1)}=f_0(P)+f_1,
$$
where $f_1$ is the linear combination of vector potentials.

We have the following from the equation (2.1)
$$
\left[ E_{1}\left(i\omega_{1}+ik_{1}v_{x}\right)\varphi_{1}+
E_{2}\left(i\omega_{2}+ik_{2}v_{x}\right)\varphi_{2}+\cdots+
E_{N}\left(i\omega_{N}+ik_{N}v_{x}\right)\varphi_{N}  \right]=
 $$
 $$
=-e \sum\limits_{j=1}^{N}\frac{E_{j}}{\omega_{j}}
\left[ k_{j}v_{y}\dfrac{\partial f_{0}}{\partial p_{x}}+
\left( \omega_{j}-k_{j}v_{x}\right)
\dfrac{\partial f_{0}}{\partial p_{y}}\right].\eqno{(\textup{2.3})}
 $$
Let us enter the dimensionless parameters
$ \Omega_{j}=\dfrac{\omega_{j}}{k_{T}v_{T}} $,
$ q_{j}=\dfrac{k_{j}}{k_{T}} $,
where $ q_{j} $ is the dimensionless wave number,
 $ k_{T}=\dfrac{mv_{_{T}}}{\hslash } $  is the heat wave number,
 $ \Omega_{j} $ is dimensionless oscillation frequency of vector
 potential electromagnetic field
$ \textbf{E}_j $.

In the equation (2.3) we will pass to the dimensionless parameters.
 We obtain the equation
$$
\left[
E_{1}\left(q_{1}P_{x}-\Omega_{1}\right)\varphi_{1}+
E_{2}\left(q_{2}P_{x}-\Omega_{2}\right)\varphi_{2}+
\cdots+E_{N}\left(q_{N}P_{x}-\Omega_{N}\right)\varphi_{N}
\right]=
$$
$$
=-e \sum\limits_{j=1}^{N}\frac{E_{j}}{\omega_{j}}\left[ q_{j}P_{y}
\dfrac{\partial f_{0}}{\partial P_{x}}+\left( \Omega_{j}-q_{j}P_{x}\right)
\dfrac{\partial f_{0}}{\partial P_{y}}\right].\eqno{(\textup{2.4})}
$$

Let us notice that  $ \dfrac{\partial f_{0}}{\partial P_{x}}\sim P_{x} $,
\qquad
$ \dfrac{\partial f_{0}}{\partial P_{y}}\sim P_{y} $.

Then
$$
\left[ q_{j}P_{y}\dfrac{\partial f_{0}}{\partial P_{x}}+
\left( \Omega_{j}-q_{j}P_{x}\right)
\dfrac{\partial f_{0}}{\partial P_{y}}\right]=
\Omega_{j}\dfrac{\partial f_{0}}{\partial P_{y}}.
$$
Now the equation (2.4) is somewhat simplified
$$
\left[
E_{1}\left(q_{1}P_{x}-\Omega_{1}\right)\varphi_{1}+
E_{2}\left(q_{2}P_{x}-\Omega_{2}\right)\varphi_{2}+
\cdots+E_{N}\left(q_{N}P_{x}-\Omega_{N}\right)\varphi_{N}
\right]=
$$
$$
=-\dfrac{e}{k_{T}p_{T}v_{T}}\sum\limits_{j=1}^{N}
E_{j}\dfrac{\partial f_{0}}{\partial P_{y}}.
\eqno{(\textup{2.5})}
$$
From the equation (2.5) we find
$$
\varphi_{j}=\dfrac{ie}{k_{T}p_{T}v_{T}}\cdot
\dfrac{\partial f_{0}/\partial P_{y}}{q_{j}P_{x}-\Omega_{j}},
\qquad j=1,2,\cdots,N.
\eqno{(\textup{2.6})}
$$
Now from the equation (2.6) we find
$$
f_{1}=\dfrac{e}{k_{T}p_{T}v_{T}}
\cdot\sum\limits_{j=1}^{N}
\dfrac{E_{j}}{q_{j}P_{x}-\Omega_{j}}.\eqno{(\textup{2.7})}
$$
Thus first approximation is defined by equality (2.7).

\begin{center}
\item{}\section{The solution of Vlasov equation in second approximation}
\end{center}

In the second approach we search for the decision of Vlasov
equation (1.2) in the form of (1.3) in which $f_2$  is defined by
equality (1.5). We substitute (1.5) in the left-hand member of equation (2.2).
We receive the following equation
$$
\sum\limits_{j=1}^{N}E_{j}^{2}\left(-2i\omega_{j}+
2ik_{j}v_{x}\right)\psi_{j}+
$$
$$
+\sum\limits_{\substack{b,s=1\\b<s}}^NE_{b}E_{s}
\left( -i(\omega_{b}+\omega_{s})+i(k_{b}+k_{s})v_{x}\right)\xi_{b,s}=
$$
$$
=-e \sum\limits_{j=1}^{N}\frac{E_{j}}{\omega_{j}}
\left[ k_{j}v_{y}\dfrac{\partial f_{0}}{\partial p_{x}}+
\left( \omega_{j}-k_{j}v_{x}\right) \dfrac{\partial f_{0}}{\partial p_{y}}\right].
\eqno{(\textup{3.1})}
$$

Let us pass in this equation to the dimensionless parameters and we will enter the
following designations

$$
q_{bs}=\dfrac{q_{b}+q_{s}}{2},\qquad
\Omega_{bs}=\dfrac{\Omega_{b}+\Omega_{s}}{2}.
$$

We receive the equation

$$
\sum\limits_{j=1}^{N}E_{j}^{2}
\left(q_{j}P_{x}-\Omega_{j}\right)\psi_{j}+
\sum\limits_{\substack{b,s=1\\b<s}}^NE_{b}E_{s}
\left(
qP_{x}-\Omega\right)\xi_{b,s}=
$$
$$
=-\dfrac{e^{2}}{2k_{T}^{2}p^{2}_{T}v_{T}^{2}}
\left\lbrace
\sum\limits_{j=1}^{N}\dfrac{E^{2}_{j}}{\Omega_{j}}
\left[ q_{j}P_{y}\dfrac{\partial}{\partial P_{x}}
\left(
\dfrac{\partial f_{0}/\partial P_{y}}{q_{j}P_{x}-\Omega_{j}}
\right)-\dfrac{\partial^{2}f_{0}}{\partial P_{y}^{2}}
\right]+\right.
$$
$$
\left.
+
\sum\limits_{\substack{b,s=1\\b<s}}^N
\dfrac{E_{b}E_{s}}{\Omega_{b}}
\left[ q_{b}P_{y}\dfrac{\partial}{\partial P_{x}}
\left(
\dfrac{\partial f_{0}/\partial P_{y}}{q_{s}P_{x}-\Omega_{s}}
\right)+\dfrac{\Omega_{b}-q_{b}P_{x}}{q_{s}P_{x}-\Omega_{s}}
\dfrac{\partial^{2}f_{0}}{\partial P_{y}^{2}}
\right]+\right.
$$
$$
\left.+\sum\limits_{\substack{b,s=1\\b<s}}^N
\dfrac{E_{s}E_{b}}{\Omega_{s}}
\left[q_{s}P_{y}\dfrac{\partial}{\partial P_{x}}
\left(
\dfrac{\partial f_{0}/\partial P_{y}}{q_{b}P_{x}-\Omega_{b}}
\right)+\dfrac{\Omega_{s}-q_{s}P_{x}}{q_{b}P_{x}-\Omega_{b}}
\dfrac{\partial^{2}f_{0}}{\partial P_{y}^{2}}
\right]\right\rbrace.
$$

 We  find from this equation
$$
\psi_{j}=-\dfrac{e^{2}}{2k^{2}_{T}p_{T}^{2}v^{2}_{T}\Omega_{j}}
\dfrac{\Xi_{j}(\textbf{P})}{q_{j}P_{x}-\Omega_{j}},
\qquad (j=1,2,\cdots,N)
\eqno{(\textup{3.2})}
$$
and
$$
\xi_{b,s}=-\dfrac{e^{2}}{2k_{T}^{2}p_{T}^{2}v_{T}^{2}}
\left[
\dfrac{1}{\Omega_{b}}\dfrac{\Xi_{bs}(\textbf{P})}{qP_{x}-\Omega}+
%\right.
%$$
%$$
%\left.
+\dfrac{1}{\Omega_{s}}\dfrac{\Xi_{sb}(\textbf{P})}{qP_{x}-\Omega}
\right],
\eqno{(\textup{3.3})}
$$
where
$$ b<s,\qquad\qquad j=1,2,\cdots,N.
$$
and
$$
\Xi_{j}(\textbf{P})=q_{j}P_{y}\dfrac{\partial}{\partial P_{x}}
\left( \dfrac{\partial f_{0}/\partial P_{y}}{q_{j}P_{x}-\Omega_{j}}
\right)-\dfrac{\partial^{2} f_{0}}{\partial P_{y}^{2}},\hspace{2cm}
%\qquad (j=1,2,\cdots,N).
$$
$$
\Xi_{bs}(\textbf{P})=q_{b}P_{y}\dfrac{\partial }{\partial P_{x}}
\left(
\dfrac{\partial f_{0}/\partial P_{y}}{q_{s}P_{x}-\Omega_{s}}\right)
+
\dfrac{\Omega_{b}-q_{b}P_{x}}{q_{s}P_{x}-\Omega_{s}}
\dfrac{\partial^{2} f_{0}}{\partial P_{y}^{2}},
$$
$$
\Xi_{sb}(\textbf{P})=q_{s}P_{y}\dfrac{\partial }{\partial P_{x}}
\left(
\dfrac{\partial f_{0}/\partial P_{y}}{q_{b}P_{x}-\Omega_{b}}\right)
+
\dfrac{\Omega_{s}-q_{s}P_{x}}{q_{b}P_{x}-\Omega_{b}}
\dfrac{\partial^{2} f_{0}}{\partial P_{y}^{2}},
$$
where
$$
b<s,\qquad b,s=1,2,\cdots,N.
$$

Thus the decision of Wigner equation is constructed and in the
second approach. It is defined by equalities (1.5) and (3.2)--(3.3). \bigskip

\section{Density of transversal electric current}

The density of electric current according his definition is equal
$$
\textbf{j}=e\int \textbf{v}f\dfrac{2d^{3}p}{\left(2\pi \hslash
    \right)^{3} } =\dfrac{2 p_{T}^{2}v_{T}}{\left(2\pi \hslash \right)^{3} }
\int f \textbf{P} d^{3}P.\eqno{(\textup{4.1})}
$$

The vector of a current density has two nonzero components
$ \textbf{j}=\left(j_{x},j_{y},0 \right)  $, where  $ j_{x} $
is density of transversal current,
 $ j_{y} $ is density of longitudinal current.

Let us calculate density of transversal current.
It is defined by the following expression
$$
j_{y}=e\int v_{y}f\dfrac{2d^{3}P}{\left( 2\pi\hslash \right)^{3} }=
e \int v_{y}f_{1}\dfrac{2d^{3}p}{\left(2\pi\hslash \right)^{3} }=
\dfrac{2p_{T}^{2}v_{T}}{\left(2\pi\hslash \right)^{3} }
\int f_{1}P_{y}d^{3}P.\eqno{(\textup{4.2})}
$$

Transversal current is directed along an electromagnetic field.
Its  density is defined according to (4.2) only first approximation of
a cumulative distribution function. The second
approximation of a cumulative distribution function does not make a
contribution to a current density.
Thus, in an explicit form transversal current is equal

$$
j_{y}=\dfrac{2ie^{2}p_{T}^{2}}{\left( 2\pi \hslash \right)^{3}k_{T} }
\int \sum\limits_{j=1}^{N} \dfrac{E_{j}}{q_{j}P_{x}-\Omega_{j}}
\dfrac{\partial f_{0}}{\partial P_{y}}P_{y}d^{3}P.\eqno{(\textup{4.3})}
$$

We  simplify a formula (4.3)
$$
j_{y}=\dfrac{2ie^{2}p_{T}^{2}}{\left( 2\pi \hslash \right)^{3}k_{T} }
\int\limits_{-\infty}^{\infty}
 \sum\limits_{j=1}^{N} \dfrac{E_{j}}{q_{j}P_{x}-\Omega_{j}}
\ln(1+e^{\alpha-P_{x}^{2}})dP_{x}.\eqno{(\textup{4.4})}
$$

\section{Density of longitudinal electric current}

We will investigate longitudinal current. By means of decomposition (1.5)
we will present longitudinal current in the following form

$$
j_x=\sum\limits_{a=1}^{N}j_{a}+
\sum\limits_{\substack{b,s=1\\b<s}}^{N}j_{bs}+
\sum\limits_{\substack{b,s=1\\b<s}}^{N}j_{sb},
\eqno{(5.1)}
$$
where
$$
j_{a}=
\dfrac{e^{3}p_{T}E_{a}^{2}}{(2\pi\hbar)^{3}k_{T}^{2}v_{T}
    \Omega_{a}}\int
\dfrac{\Xi_{a}(\textbf{P})P_{x}d^{3}P}{q_{a}P_{x}-\Omega_{a}},
\quad (a=1,2,\cdots,N),\eqno{(5.2)}
$$
and
$$
j_{bs}=\dfrac{e^{3}p_{T}}{(2\pi\hbar)^{3}k_{T}^{2}v_{T}}
\int
\dfrac{E_{b}E_{s}}{\Omega_{b}}
\dfrac{\Xi_{bs}\textbf{P}}{qP_{x}-\Omega}P_{x}d^{3}P,
\eqno{(5.3)}
$$

$$
j_{sb}=\dfrac{e^{3}p_{T}}{(2\pi\hbar)^{3}k_{T}^{2}v_{T}}
\int
\dfrac{E_{b}E_{s}}{\Omega_{s}}
\dfrac{\Xi_{sb}\textbf{P}}{qP_{x}-\Omega}
 P_{x}d^{3}P.
\eqno{(5.4)}
$$
Here
$$
q_{bs}=\dfrac{q_{b}+q_{s}}{2}, \qquad
\Omega_{bs}=\dfrac{\Omega_{b}+\Omega_{s}}{2} ,\quad b<s,\quad b,s=1,2,\cdots,N.
$$

In these expressions one-dimensional internal integral on
$ P_{y} $ is equal to zero and internal integral for $ P_ {x} $
are calculated piecemeal. Therefore, the previous equalities
 becomes simpler for components of longitudinal current. Then,
 internal integral we will integrate on a variable of $ P_ {y} $.
 Further we will calculate internal integrals in plane $ (P_{y},P_{z}) $
 in polar coordinates. Equalities (5.2) -- (5.4) come down to one-dimensional integral.
  $$
  j_{a}=\dfrac{\pi e^{3}p_{T}E_{a}^{2}q_{a}}{(2\pi\hbar)^{3}k_{T}^{2}v_{T}}
  \int\limits_{-\infty}^{\infty}
  \dfrac{\ln\left(1+e^{\alpha-P_{x}^{2}} \right)
    dP_{x} }{\left( q_{a}P_{x}-\Omega_{a}\right)^{3} },\qquad (a=1,2,\cdots,N).
  $$
 and
 $$
 j_{bs}=\dfrac{\pi e^{3}p_{T}E_{b}E_{s}q_{b}\Omega}{(2\pi\hbar)^{3}
    k_{T}^{2}v_{T}\Omega_{b}}
 \int\limits_{-\infty}^{\infty}
 \dfrac{\ln\left( 1+e^{\alpha-P_{x}^{2}} \right)
    dP_{x} }{\left(q_{s}P_{x}-\Omega_{s} \right)
 \left( qP_{x}-\Omega\right)^{2}},
 $$
 $$
 j_{sb}=\dfrac{\pi e^{3}p_{T}E_{b}E_{s}q_{s}\Omega}{(2\pi\hbar)^{3}
    k_{T}^{2}v_{T}\Omega_{s}}
 \int\limits_{-\infty}^{\infty}
 \dfrac{\ln\left( 1+e^{\alpha-P_{x}^{2}} \right)
    dP_{x} }{\left(q_{b}P_{x}-\Omega_{b} \right)
    \left( qP_{x}-\Omega\right)^{2}},
 $$
 $ \quad b<s,\qquad b,s=1,2,\cdots,N. $

 Let us find numerical density the concentration of particles of
  plasma answering to distribution of Fermi---Dirac
 $$
 N_{0}=\int f_{0}(P)\dfrac{2d^{3}p}{\left( 2\pi\hslash\right)^{3} }=
 \dfrac{8\pi p_{T}^{3}}{\left( 2\pi\hslash \right)^{3} }
  \int\limits_{0}^{\infty}
 \dfrac{e^{\alpha-P^{2}}P^{2}dP}{1+e^{\alpha-P^{2}}}=
 \dfrac{k_{T}^{3}}{2\pi^{2}}l_{0}(\alpha),
 $$
 where
 $$
 l_{0}(\alpha)= \int\limits_{0}^{\infty}
  \ln\left( 1+e^{\alpha-\tau^{2}}\right)d\tau.
 $$

 We will enter plasma (Langmuir) frequency in expression before integrals
  $$  \omega_{p}=\sqrt{\dfrac{4\pi e^{2}N_{0}}{m}} $$
  and numerical density (concentration)
    $N_{0} $. We will express numerical density through a thermal wave number.
    Then
 $$
 \dfrac{\pi p_{T} e^{3}q_{j}}{\left( 2\pi\hslash \right)^{3}
    k_{T}^{2} v_{T} }=\dfrac{e \Omega_{p}^{2}}{p_{T}k_{T}}\cdot
 \dfrac{k_{j}}{16\pi l_{0}(\alpha)}
 =\sigma_{\rm l,tr}\dfrac{k_{j}}{16\pi l_{0}(\alpha)},
 \qquad (j=1,2,\cdots,N).
 $$
 Here
$$
\Omega_{p}=\dfrac{\omega_{p}}{k_{T}v_{T}}=
 \dfrac{\hslash \omega_{p}}{mv_{T}^{2}}
$$
is the dimensionless  plasma (Langmuier) frequency,
 $ \sigma_{\rm l,tr} $ is the longitudinal-transversal conductivity,
$$
\sigma_{\rm l,tr}=\dfrac{e\Omega_{p}^{2}}{p_{T}k_{T}}.
$$

 Now we will write down components of longitudinal current in form

 $$
 j_{a}=E_{a}^{2}\sigma_{\rm l,tr}k_{a}J_{a},\quad
 j_{bs}=E_{b}E_{s}\sigma_{\rm l,tr} k_{b}J_{bs},\quad
 j_{sb}=E_{b}E_{s}\sigma_{\rm l,tr} k_{s}J_{sb},
 \eqno{(5.5)}
 $$
 where
 $$
 J_{a}=\dfrac{1}{16\pi l_{0}(\alpha)}
 \int\limits_{-\infty}^{\infty}
 \dfrac{\ln\left(1+e^{\alpha-P_{x}^{2}}\right)dP_{x} }
 {\left( q_{a}P_{x}-\Omega_{a} \right)^{3} },\qquad (a=1,2,\cdots,N),
 $$
 $$
 J_{bs}=\dfrac{\Omega}{16\pi l_{0}(\alpha)\Omega_{b}}
 \int\limits_{-\infty}^{\infty}
 \dfrac{\ln\left(1+e^{\alpha-P_{x}^{2}}\right)dP_{x} }
 {\left( q_{s}P_{x}-\Omega_{s} \right)
    \left(qP_{x}-\Omega\right)^{2} },\quad (b<s,\qquad b,s=1,2,\cdots,N),
 $$
 $$
 J_{sb}=\dfrac{\Omega}{16\pi l_{0}(\alpha)\Omega_{s}}
 \int\limits_{-\infty}^{\infty}
 \dfrac{\ln\left(1+e^{\alpha-P_{x}^{2}}\right)dP_{x} }
 {\left( q_{b}P_{x}-\Omega_{b} \right)
    \left(qP_{x}-\Omega\right)^{2} },\quad (b<s,\qquad b,s=1,2,\cdots,N).
 $$
Here
$$
 q=\dfrac{q_{b}+q_{s}}{2},\qquad  \Omega=\dfrac{\Omega_{b}+\Omega_{s}}{2}.
$$

 In equalities (5.5)
 $ J_{1},J_{2},\cdots,J_{N},J_{12},J_{21},\cdots,J_{bs},J_{sb} $
 are the dimensionless parts of density of longitudinal current.
Thus, a longitudinal part of current is equal
 $$
 j_{x}=\sigma_{\rm l,tr}
 \left[
 \sum\limits_{a=1}^{N}E_{a}^{2}k_{a}J_{a}+
 \sum\limits_{\substack{b,s=1\\b<s}}^{N} E_{b}E_{s}
 \left( k_{b}J_{bs}+k_{s}J_{sb}\right)
\right].\eqno(5.6)
 $$
If to enter transversal fields
 $$
 \textbf{E}_{j}^{\rm tr}=\textbf{E}_{j}-
 \dfrac{\textbf{k}_{j}(\textbf{E}_{j}\textbf{k}_{j})}{k_{j}^{2}},
 $$
 then equality (5.6) can be written down in an invarianty form
 $$
 \textbf{j}^{\rm long}=\sigma_{\rm l,tr}
  \left[
 \sum\limits_{a=1}^{N}(\textbf{E}_{a}^{\rm tr})^{2}\textbf{k}_{a}J_{a}+
 \sum\limits_{\substack{b,s=1\\b<s}}^{N} \textbf{E}_{b}^{\rm tr}
\textbf{E}_{s}^{\rm tr}
 \left( \textbf{k}_{b}J_{bs}+\textbf{k}_{s}J_{sb}\right)
 \right].
 $$
Let us consider a case of small values of a wave number. From
(5.6) follows that at small values of wave numbers for density of
 longitudinal current we receive
 $$
 j_{x}=-\dfrac{\sigma_{\rm l,tr}}{8\pi}
 \left[
 \sum\limits_{a=1}^{N}E_{a}^{2}\dfrac{k_{a}}{\Omega_{a}^{3}}+
 2\sum\limits_{\substack{b,s=1\\b<s}}^{N} E_{b}E_{s}
 \dfrac{k_{b}+k_{s}}{\Omega_{b}\Omega_{s}
    \left( \Omega_{b}+\Omega_{s} \right) }
   \right].\eqno(5.7)
 $$

 \textbf{Remark}

 At calculation of the integrals entering the dimensionless parts of density
  of longitudinal current it is necessary to use Landau's rule. According to
   this rule, for example, for integrals  $ J_{a} $ we will have
$$
 J_{a}=\dfrac{1}{16\pi l_{0}(\alpha)}
 \Bigg[-\dfrac{i\pi}{2q^{3}_{a}}\Big[ \ln\Big( 1+e^{\alpha-\tau^{2}} \Big)
 \Big]^{''}_{r=\frac{\Omega_{a}}{q_a}}
+{\rm V.p.}\int\limits_{-\infty}^{\infty}\dfrac{\ln \Big( 1+e^{\alpha-\tau^{2}}\Big)
d\tau }{(q_{a}\tau-\Omega_{a})^{3}}\Bigg],
$$
where  $a=1,2,\cdots,N.$

\section{Conclusions}

We found out dependence of transversal and longitudinal current,
 generated in classical plasma by $ N $  transverse electromagnetic waves.
In this work we consider the effect of the nonlinear character of
 the interaction of electromagnetic fields with collisionless
  Maxwell classical plasma.
 We considered the Vlasov equation and his solution, the method
 of successive approximations has been found of the
 distributions functions.
 We found formulas for electric current in collisionless classical plasma.

\end{document}